# Field Enhancement in Plasmonic Nanostructures

Shiva Piltan[1,*] and Dan Sievenpiper[1,*]

[1]Electrical and Computer Engineering Department, University of California San Diego, La Jolla, California 92093, USA

[*]spiltan@eng.ucsd.edu; dsievenpiper@eng.ucsd.edu.

Efficient generation of charge carriers from a metallic surface is a critical challenge in a wide variety of applications including vacuum microelectronics and photo-electrochemical devices. Replacing semiconductors with vacuum/gas as the medium of electron transport offers superior speed, power, and robustness to radiation and temperature. We propose a metallic resonant surface combining optical and electrical excitations of electrons and significantly reducing powers required using plasmon-induced enhancement of confined electric field. The properties of the device are modeled using the exact solution of the time-dependent Schrödinger equation at the barrier. Measurement results exhibit strong agreement with an analytical solution, and allow us to extract the field enhancement factor at the surface. Significant photocurrents are observed using combination of $\frac{W}{cm^2}$ optical power and 10 volts DC excitation on the surface. The model suggests optical field enhancement of 3 orders of magnitude at the metal interface due to plasmonic resonance. This simple planar structure provides valuable evidence on the electron emission mechanisms involved and it can be used for implementation of semiconductor compatible vacuum devices.

The coupling of electromagnetic radiation to free electron oscillations at a metal interface and its consequent properties including enhanced optical near-field have drawn significant interest to the field of plasmonics and its applications. Some of the most widespread demonstrated applications of this nanoscale light-matter interaction include surface-enhanced Raman spectroscopy (SERS),[1,2] plasmonic color pixels for CMOS compatible imaging and bio-sensing,[3-6] hot electron photo-electrochemical and photovoltaic devices and photodetectors,[7-10] optical antennas,[11,12] plasmonic integrated circuits,[13,14] and metamaterials.[15,16] Here, we propose a simple periodic surface (Fig. 1) to experimentally quantify the enhancement of linear and nonlinear optical processes involved in this nanoscale light manipulation. The experimental results are in strong agreement with theoretical analysis which is done using the exact solution of the time-dependent Schrödinger equation at the barrier and subwavelength enhancement of electromagnetic fields on the order of 1000 are achieved in a simple planar structure. The observed enhancement can be exploited in applications requiring efficient generation of charge carriers such as photovoltaics and plasmon-mediated photochemistry by replacing metal as the active component.[17]

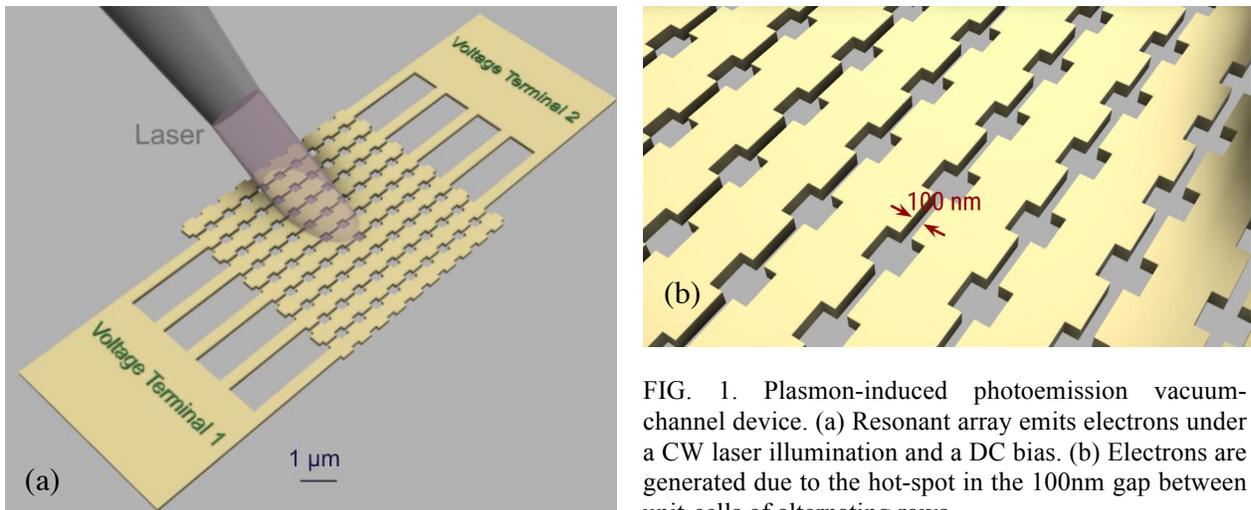

FIG. 1. Plasmon-induced photoemission vacuum-channel device. (a) Resonant array emits electrons under a CW laser illumination and a DC bias. (b) Electrons are generated due to the hot-spot in the 100nm gap between unit-cells of alternating rows.

Following the discovery of the first solid-state diode in 1874[18] and the phenomenon behind thermionic emission,[19] the first practical vacuum tube device, the thermionic diode, was patented in 1904.[20] Although solid state devices provide superior life-time and integrability[21], vacuum still performs better as a medium for ballistic charge transport enabling speeds up to 3 orders of magnitude faster[22,23]. Also, advanced fabrication techniques can be used to reduce gap size to a fraction of mean free path length of an electron moderating cathode degradation and high vacuum requirements. However, liberating electrons from the metal surface efficiently remains a challenging task. Our proposed plasmonic resonant surface provides the means for combining optical and electrical generation of electrons. Taking advantage of the confined field enhancement, it relaxes both optical field (from $\frac{GW}{cm^2}$ in mid-IR[24] to $\frac{W}{cm^2}$) and bias voltage (from 100s of volts[25,26] to less than 10V) requirements.

DC and laser induced electron emission processes can be categorized in four main mechanisms as shown in Fig. 2, field emission, photo-assisted field emission, photoemission, and optical field emission.

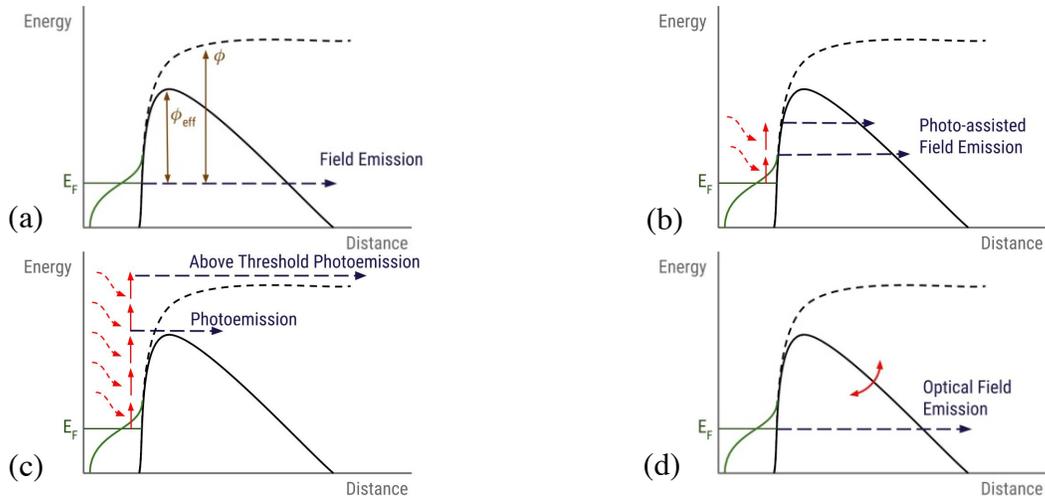

FIG. 2. Electron emission mechanisms under DC and optical illumination. (a) Field emission, effective work function is reduced due to the image charge effect. (b) Photo-assisted field emission, electron energy is enhanced by one/two photon energy and tunnels subsequently. (c) Photoemission, electron absorbs sufficient photons to travel over the barrier, and above threshold photoemission, electron is emitted above the barrier by absorbing more photons than required. (d) Optical field emission, the barrier is modulated due to the optical excitation and at parts of the optical cycle it is narrow enough for electrons to tunnel directly from the Fermi level.

The barrier at electric field strengths above $1\frac{V}{nm}$ becomes thin enough for the electrons to tunnel through from states close to Fermi level. This process [Fig. 2(a)] is referred to as field emission and was analytically characterized by Fowler and Nordheim by solving the time-independent Schrödinger equation for a rounded triangular barrier.[27] The barrier is rounded due to the image charge effect and has a lower effective work function.[28,29] The total current density can be obtained as a function of electric field at the surface, electron temperature, and Fermi energy for electrons with energy $\varepsilon$ in the surface normal direction.[22,29,30,31] It can be concluded that for the strong-field tunneling mechanism, $\ln(\frac{I}{V^2})$ versus $\frac{1}{V}$ curves should be linear with a negative slope depending on the work function and the surface field enhancement.

The addition of optical excitation can interact with the system by both modulating the barrier and increasing the initial energy of electrons due to photon absorption.[29,32-34] The dominant mechanism for weak optical fields is either photo-assisted field emission [Fig. 2(b)] or photoemission [Fig. 2(c)]. In the photo-assisted field emission process, electron energy is enhanced to a non-equilibrium distribution by absorption of one/two photons of frequency $w$ and tunnels through thereafter. In the photoemission process, the electron absorbs a sufficient number of photons to travel above the barrier as shown in Fig. 2(c). If the multiphoton emission process happens by absorption of a greater number of photons than required ($n > \phi/\hbar w$) the process is called above-threshold photoemission.[35,36] It has been shown than in the multiphoton regime photo-emitted current is proportional to the nth power of laser intensity, where n is the total number of photons absorbed.[30,35]

If the laser intensity is sufficiently large, the potential barrier becomes narrow enough during part of the optical cycle for the electrons to tunnel through directly from the Fermi level. This process is called optical field emission and it is shown in Fig. 2(d). Fowler-Nordheim type equations can be used in this strong-field regime by substituting the total field with the sum of DC and AC fields.

The transition between the photon-driven and field-driven regimes is characterized by Keldysh parameter which is defined as the ratio of the incident optical frequency to the characteristic tunneling frequency of metal.[37,38] For relatively weak fields where Keldysh parameter is above unity, the quiver amplitude of electrons in the applied field is smaller than the decay length of the near field; therefore, electrons are back accelerated before they are liberated and the dominant process is multiphoton induced emission.[24,32] The local slope of the photocurrent as a function of laser intensity drops as the dominant mechanism changes from multi-photon induced emission to optical field emission and therefore it is an appropriate characterizing property.[24,38]

Nanostructures supporting surface plasmon resonances provide both electric field enhancement and spatial confinement. The induced local field enhancement can be used to facilitate electron emission and ponderomotive acceleration processes and it also provides access to high field intensity regions by operating in intensities below the damage threshold of the metal.[24,30,35,36,38-42] The resonant field enhancement is due to the collective oscillation of conduction band electrons in a confined region and it depends on the geometrical shape; therefore, it provides an additional control parameter to liberate electrons.[40,43,44]

Here, we have optimized a resonant metallic array of gold unit-cells to maximize the field enhancement when illuminated at 785nm wavelength. The consequent local field enhancement enables the unit-cells to emit electrons at DC and optical power levels significantly below expected. The array consists of multiple rows of unit-cells as shown in Fig. 3(a). There are two terminals serving as the feed for the DC excitation and every other row of the array is connected to one terminal. Therefore, a vacuum channel is formed between the plates of each unit-cell on two adjacent rows.

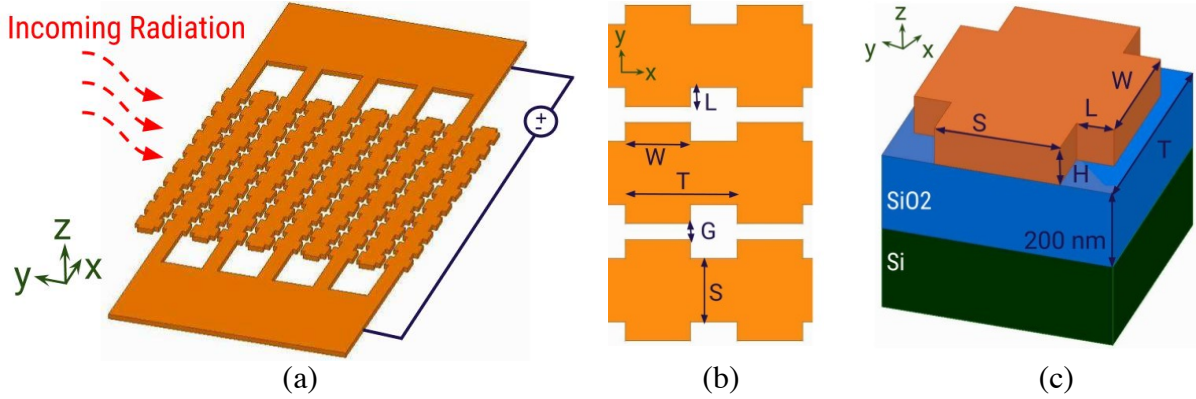

FIG. 3. Schematic of the resonant vacuum-channel device. (a) DC biased array under laser illumination. (b) Top-view of multiple cells, W= 450nm, L= 120nm, T= 770nm, S= 400nm, H= 140nm, and G= 100nm. (c) Simulation unit-cell, 140nm of e-beam evaporated gold on top of 200nm isolating $SiO_2$ on Si wafer.

The dimensions of the geometry, including the gap between the plates G, the period of unit-cells T, the width and length of the fingers W and L, and the thickness of gold layer H have been optimized to provide the maximum electric field enhancement at the center point between the plates of unit-cells on two adjacent rows upon excitation with a plane wave at 785nm, polarized in the y-direction. The design simulations were done using HFSS electromagnetics tool based on the finite-element method, and using the Johnson-Christy model for the complex dielectric constant properties of gold.[45] Field enhancement is defined as the ratio of the electric field in the middle point in the gap to the incident field. Fig. 4(a) demonstrates the simulated field enhancement as a function of wavelength. It is shown that the electric field in the gap is enhanced approximately by a factor of 10 at the resonant mode in 785nm.

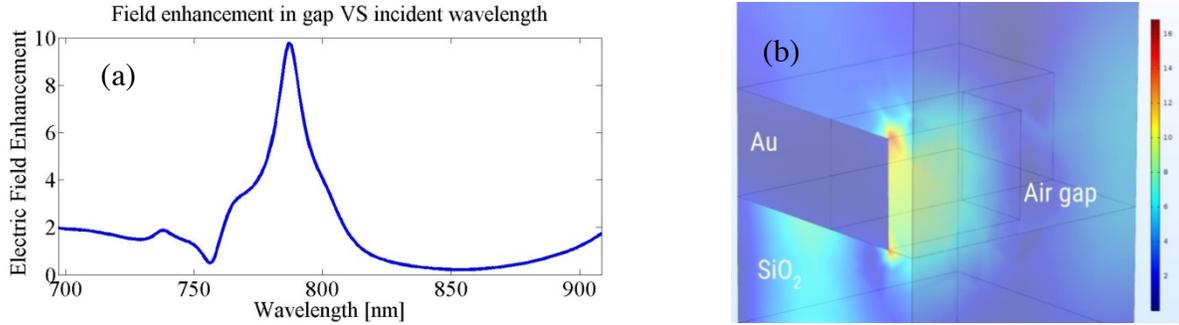

FIG. 4. (a) Electric field enhancement in the gap as a function of incident wavelength. (b) Simulated field profile in half unit-cell at 785nm, with the maximum field confined in the air gap.

The designed array was fabricated on silicon wafers with a 200nm thermally oxide $SiO_2$ layer for isolation. A 40x30 element layout was exposed using electron beam lithography technique. A 10nm chromium adhesion layer and 130nm of gold were e-beam evaporated subsequently. The fabricated arrays were wire-bonded in standard dual in-line packages. Fig. 5 shows SEM images of fabricated samples.

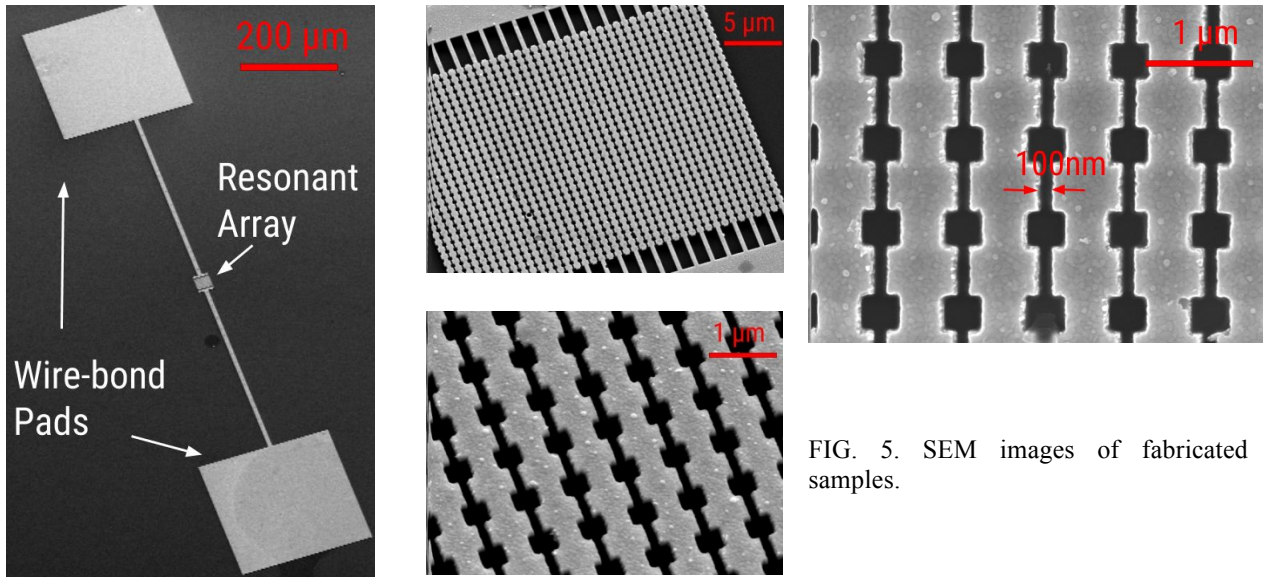

FIG. 5. SEM images of fabricated samples.

The average field enhancement factor in fabricated samples was first confirmed using Raman spectroscopy to be approximately 10. The measurement setup consists of a tunable Ti:Sapphire CW laser pumped with a 10 watt green semiconductor laser. The beam was focused on the sample inside a vacuum chamber pumped down to 0.1 mTorr. The laser beam was sampled for power and wavelength measurements outside vacuum chamber using a silicon photo-detector and spectrometer. The measured I-V curves of the device are shown in Fig. 6(a) and 6(b). The laser intensity was changed from 0 to 140 $\frac{W}{cm^2}$ keeping the wavelength at 785 nm. For each fixed laser power the DC voltage was swept from -12 to 12v resulting in an electric field in the gap ranging from 0 to 0.12 $\frac{v}{nm}$. For low laser power the DC electric field is not sufficient for the electrons to overcome the potential barrier as expected; however, it is shown that the emitted current increases significantly as the optical illumination intensifies to levels as low as 10s of $\frac{W}{cm^2}$. This is due to the highly nonlinear light-matter interaction at the surface allowing for a combination of DC and AC field excitations. If the electrons are mainly liberated due to n-photon absorption mechanism, $I \propto P^n$ where I is the photocurrent and P is laser power; therefore, I-P curves in logarithmic scale are linear with a slope showing the number of photons being absorbed. We have shown the I-P curves in linear [Fig. 6(c)] and logarithmic [Fig. 6(d)] scales for different measurements. The slope of I-P curves in log scale is between 1 and 2 in our measurements, indicating the dominance of one/two photon absorption mechanism.

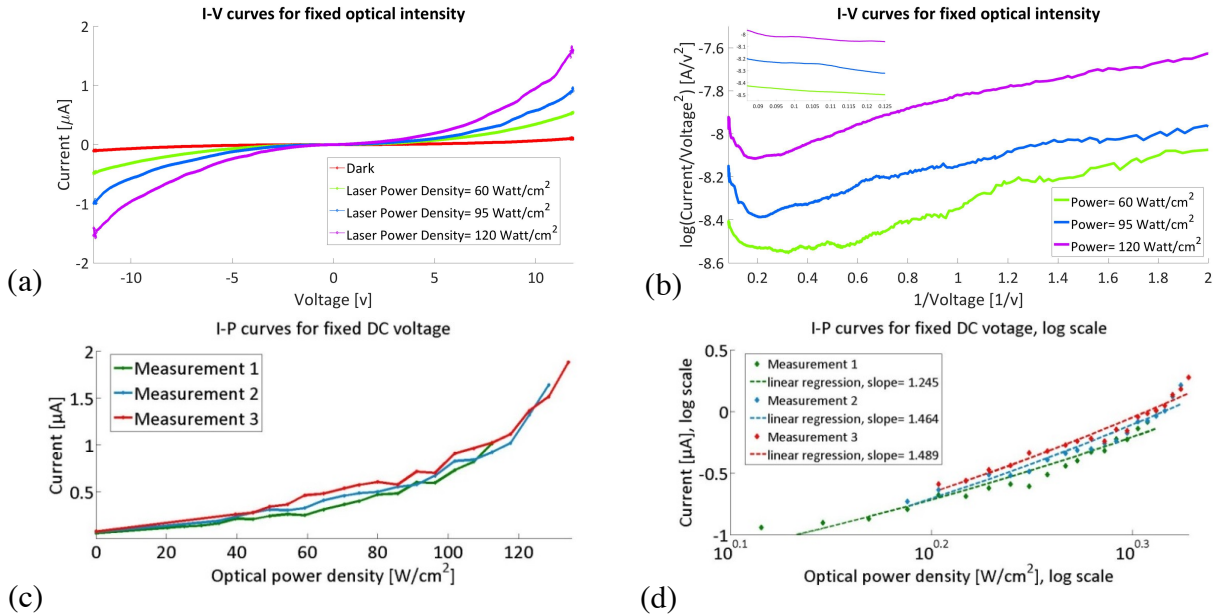

FIG. 6. (a) Measured current as a function of DC voltage, for fixed laser power intensities (0, 60, 95, and 120 $\frac{W}{cm^2}$ shown). (b) $\log(I/V^2)$-$V^{-1}$ curves for optical intensity of 60, 95, and 120 $\frac{W}{cm^2}$, inset for DC voltage above 8 v. (c) Current versus optical power density for fixed DC electric field (0.11 $\frac{v}{nm}$ shown). (d) Current versus optical power density for fixed DC electric field (0.11 $\frac{v}{nm}$) in log scale and their linear regression, the slope is an indication of the number of photons being absorbed.

As discussed in the first section, optical illumination can interact with the electron emission process by both modulating the potential barrier and increasing initial energy of electrons due to photon absorption. The confined electric field enhancement induced by resonant plasmons facilitates liberation of electrons in two ways, it provides access to high-field regions and the intense field gradient due to the nanoscale confinement enhances the ponderomotive acceleration. Ward *et al.* demonstrated field enhancements exceeding 1000 in nanogaps between plasmonically active gold nanostructures.[46] They extracted the enhanced electric field and tunneling conduction of nanogaps ranging from 0.03 to 0.23 nm using a CW laser at 785 nm and peak intensity of 22.6 $\frac{kW}{cm^2}$. Dombi *et al.* demonstrated the role of plasmonic field enhancement in electron emission from resonant metallic nanoparticles and showed maximized photoemission and kinetic energies at the resonance frequency.[40] It has also been shown that the work function of aresonant nanostructures supporting plasmon coupling is effectively decreased upon interaction with incoming photons by a value corresponding to the energy of the photon at the laser wavelength.[29,31,35,36,39] Thermally enhanced field emission process is not of significant contribution for static fields below on the order of 1 $G\frac{v}{m}$ as discussed by Kealhofer *et al.* for hafnium carbide tips[33]. The maximum static field in our measurements is an order of magnitude lower than that limit and the optical field does not exceed 22.4 $\frac{kv}{m}$. Fig. 6(b) also shows the Fowler-Nordheim type I-V characteristics and linear behavior in higher voltages confirming the fact that emitted electrons are mainly traveling through the free space. The 200nm SiO₂ layer provides sufficient isolation so that silicon absorption would not have a significant effect as Forati *et al*. experimentally demonstrated.[47]

In order to model the device we have utilized the exact solution of the time-dependent Schrödinger equation at a modulated triangular potential barrier discussed by Zhang and Lau.[48] The model solves the Schrödinger equation for the complex electron wave function assuming zero potential inside the metal. The potential outside metal is taken as a constant DC-field induced triangular barrier modulated by a continuous AC laser field. It is shown that reflected and transmitted electron wave functions from the barrier consist of a series of ladder eigenstates with eigen-energies of $\varepsilon + n\hbar\omega$ corresponding to sub-bands available by n-photon absorption (n>0) or emission (n<0) where $\varepsilon$ is the surface normal energy of incident electron.[49] In our model, we used optical field enhancement $\beta$ and effective work function $\phi_{eff}$ as the control parameters to fit the measured data to analytical solution, assuming optical illumination at 785nm and electrons being initially at the Fermi level.[27,31,36,38,50] The model does not include the reduction of barrier height due to image charge effects and solves the Schrödinger equation assuming a sharp triangular barrier; however, as shown by Zhang and Lau[48] it can be used for realistic potentials including the image charge effects by replacing the work function with reduced $\phi_{eff}$. In order to fit both constant DC field and constant AC field curves to the analytical solution, the effective work function of gold had to be reduced to 3.3ev. This reduction is slightly higher than the energy of a photon at 785nm (1.58ev), suggesting a combination of one and two photon absorption at the surface. The optical field at the surface had to be enhanced by 3 orders of magnitude in order to reach measured current levels. Fig 7 shows that the analytical I-V characteristic of the device falls within the range of measured data for each optical intensity.

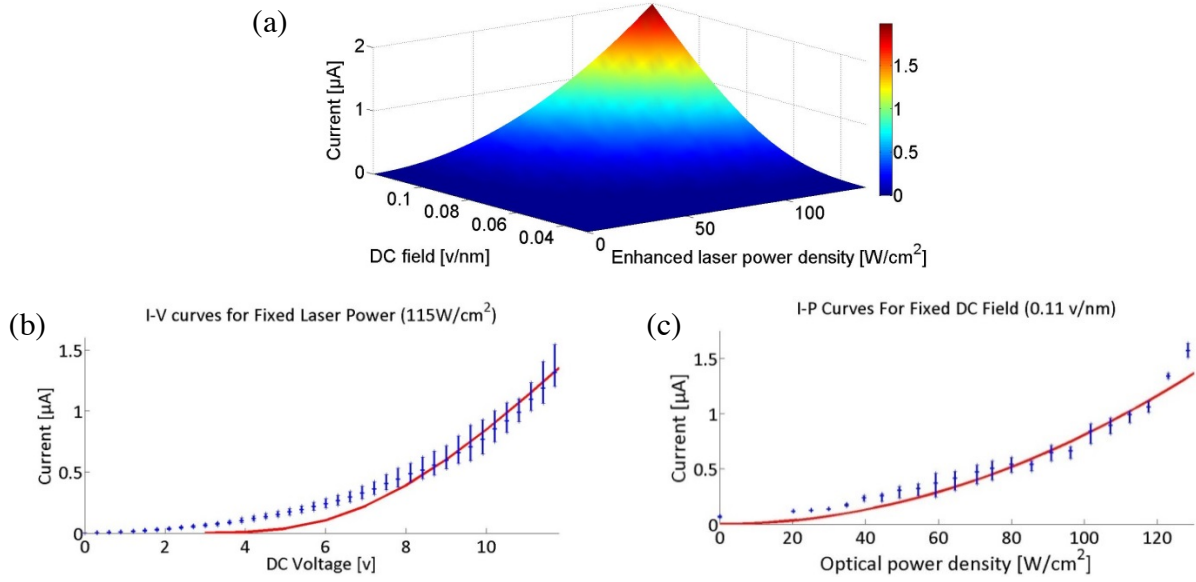

FIG. 7. Comparison between analytical model and measurement results. (a) Total time averaged current as a function of DC field and enhanced laser power density based on analytical solution. (b) I-V curves at laser intensity= $115 \frac{W}{cm^2}$, red: analytical model, blue: measured data. (C) I-P curves at DC field= $0.11 \frac{v}{nm}$, red: analytical model, blue: measured data.

Fig. 7(a) shows the effect of combination of DC and AC excitation in electron emission. The maximum laser intensity even after applying the enhancement factor in the model is below $0.04 \frac{v}{nm}$ which is considerably smaller than the DC field to have a significant effect on the current; however, due to the nonlinear properties of the laser-matter interaction, the consequent photocurrent is increased substantially. The analytical model does not include the induced field inside the metal and assumes a zero potential inside gold. The extracted surface field enhancement is also compatible with values obtained by Ward *et al*[46] at 785 nm and optical peak intensity of $22.6 \frac{kW}{cm^2}$. The surface enhancement mechanism has been previously utilized in ultrafast free electron generation.[24,31,51] It has also been discussed that photo-assisted field emission process can be favored over multiphoton over-the-barrier photoemission due to the increase in tunneling probability of electrons through an optically modified barrier.[35] Therefore, it appears from our experiments that the optical excitation at the resonance wavelength of the surface reduces the gold work function at least by one photon energy. Additionally, the enhanced optical field induces sufficient modulation of the barrier to make it narrow enough at parts of the optical cycle for the electrons to tunnel through after absorption of one or two photon energies. The obtained photocurrent based on the analytical model strongly agrees with the measured values in terms of both DC voltage and optical power levels as depicted in Fig. 7(b) and Fig. 7(c) respectively.

In summary, we have designed, fabricated, and modeled a plasmon-induced photoemission based vacuum-channel device enabling efficient combination of DC and AC field induced emission of electrons. Meanwhile neither solely optical nor DC excitation per se provides sufficient energy for the electrons to overcome the barrier, their highly nonlinear interaction at the interface results in significant photocurrent which has been characterized in our measurements. The laser and DC power levels required for electron emission are substantially reduced ($\frac{W}{cm^2}$ optical intensity and less than 10v DC bias) in a simple planar metallic surface due to the plasmon-induced subwavelength field concentration and manipulation. The effective work function of gold and the optical electric field enhancement factor at the surface have been quantified using the exact solution of the time-dependent Schrödinger equation in a field-modulated triangular barrier. It is shown that the experimental results fit well in the analytical model having an effective work function of 3.3ev and a field enhancement of 1800. Utilization of the resonant field enhancement in this periodic structure addresses the challenges of efficient generation of carriers in vacuum-channel devices and can be taken advantage of in a variety of applications including nanoscale photonic circuits, photovoltaics, and photochemistry.

The authors thank E. Forati, T. J. Dill, and UC San Diego nanofabrication facility staff including S. Parks, L. Grissom, R. Anderson, I. Harris, and X. Lu for the helpful discussions, and especially Dr. M. Montero for performing Ebeam lithography exposures. This work was funded by Defense Advanced Research Projects Agency (DARPA) through grant N00014-13-1-0618.